\documentclass[12pt]{article}
\usepackage{amsmath,amssymb}
\usepackage{graphicx}
\usepackage{wrapfig}

\usepackage{hyperref}
\usepackage{cite}

\def\beq{\begin{eqnarray}}
\def\eeq{\end{eqnarray}}

\newcommand{\p}{\partial}

\newcommand{\pl}{\left(}
\newcommand{\crr}{\right]}
\newcommand{\pr}{\right)}
\newcommand{\crl}{\left[}

\newcommand{\lb}{\lambdaslash}

\newcommand{\nn}{\nonumber}   

\newcommand{\be}{\begin{equation}}
\newcommand{\ee}{\end{equation}}
\newcommand{\bea}{\begin{eqnarray}}
\newcommand{\eea}{\end{eqnarray}}
\newcommand{\bg}{\begin{gather}}
\newcommand{\eg}{\end{gather}}
\newcommand{\bseq}{\begin{subequations}}
\newcommand{\eseq}{\end{subequations}}

\def\be{\begin{eqnarray}}
\def\ee{\end{eqnarray}}
\def\lb{\label}

\begin{document}

\title{\textbf{Dual Massive Gravity }}
\author{ \textbf{
Kevin Morand$^\star$ and  Sergey N. Solodukhin$^\sharp$ }} 

\date{May 9, 2012}
\maketitle

\begin{center}
  \hspace{-0mm}
  \emph{ Laboratoire de Math\'ematiques et Physique Th\'eorique, }\\
  \emph{Universit\'e Fran\c cois-Rabelais Tours, F\'ed\'eration Denis Poisson - CNRS, }\\
  \emph{Parc de Grandmont, 37200 Tours, France} \\
\end{center}

{\vspace{-11cm}
\begin{flushright}
\end{flushright}
\vspace{11cm}
}



\begin{abstract}
\noindent { The linearized massive gravity in three dimensions, over any maximally symmetric background, is known to be presented in a self-dual form  as a first order equation which encodes not only the massive Klein-Gordon type field equation but also the supplementary transverse-traceless conditions. We generalize this construction to higher dimensions.  
The appropriate dual description in d dimensions, additionally to a (non-symmetric) tensor field $h_{\mu\nu}$, involves an extra rank-(d-1) field equivalently represented by the torsion rank-3 tensor.  The symmetry condition for $h_{\mu\nu}$  arises on-shell as a consequence of the field equations.   The action principle of the dual theory is formulated. The focus has been made on four dimensions. 
Solving one of the fields in terms of the  other and putting back in the action one obtains two other equivalent formulations of the theory in which the action is quadratic in derivatives.  In  one of these representations the theory is formulated entirely in terms of a rank-2 non-symmetric tensor
$h_{\mu\nu}$. This quadratic  theory is not identical to the Fierz-Pauli theory and contains the coupling between the symmetric and antisymmetric parts of $h_{\mu\nu}$. Nevertheless, the only singularity in the propagator is the same as in the Fierz-Pauli theory so that only the massive spin-2 particle is propagating. In the other representation, the theory is formulated in terms of the torsion rank-3 tensor only. We analyze the conditions which follow from the
field equations and show that they restrict to 5 degrees of freedom thus producing an alternative 
description to the massive spin-2 particle. A generalization to higher dimensions is suggested.}
\end{abstract}

\vskip 2 cm
\noindent
\rule{7.7 cm}{.5 pt}\\
\noindent 
\noindent
\noindent ~~~$^{\star}$ {\footnotesize e-mail: Kevin.Morand@lmpt.univ-tours.fr}\\
\noindent ~~~$^{\sharp}$ {\footnotesize e-mail: Sergey.Solodukhin@lmpt.univ-tours.fr}\\


\newpage

\section{ Introduction}

Massive gravity is one of the interesting directions of current research actively discussed in the literature. The on-going research appears to converge towards a formulation of a consistent non-linear theory   (for a review of the current status and for the recent developments see \cite{Hinterbichler:2011tt}).
Nevertheless, it is always  desirable to develop alternative ways to approach the problem.
In the present paper we develop one such approach, at the present stage still at the linearized level, based on the use of first order field equations. 

The starting point for the present work is the observation \cite{Aragone:1986hm}, \cite{Tyutin:1997yn} that in three spacetime dimensions, on the background of a 
maximally symmetric metric, 
the wave equation for a massive graviton together with the supplementary (gauge) conditions 
can be written as a single first order equation
\be
\epsilon_\mu^{\ \alpha\beta}\nabla_\alpha h_{\beta\nu}=mh_{\mu\nu}\, ,
\lb{1}
\ee
where $\nabla_\alpha$ is a covariant derivative with respect to a maximally symmetric metric $g_{\alpha\beta}$.
Indeed, provided the tensor field $h_{\mu\nu}$ is symmetric and satisfies this equation, 
then it is automatically traceless and transverse,
\be
g^{\mu\nu}h_{\mu\nu}=0\, , \ \ \nabla^\mu h_{\mu\nu}=0\, ,
\lb{2}
\ee
and, squaring the equation (\ref{1}), we produce the second order equation
\be
(\Box-\frac{1}{2}R+(-1)^sm^2)h_{\mu\nu}=0\, ,
\lb{3}
\ee
where $\Box=\nabla^\alpha\nabla_\alpha$ and $s$ is the signature of spacetime. The equation (\ref{1})  appears in the   linearized gravitational equations (see for instance \cite{Sachs:2008gt}) obtained by varying the gravitational action which is the
sum of the Ricci scalar and the gravitational Chern-Simons term, a model of massive gravity first proposed in \cite{Deser:1982vy}.

In a generalization of the  equation (\ref{1}) to higher dimensions, assuming that in the right hand side there still should stand the rank-2 tensor $h_{\mu\nu}$, we find that
the left hand side should contain a tensor of rank-(d-1), 
\be
\epsilon_{\mu}^{\ \beta\alpha_1..\alpha_{d-2}} \nabla_\beta B_{\alpha_1..\alpha_{d-2}, \nu}=m_1h_{\mu\nu}
\lb{4}
\ee
Thus, in higher dimensions we would need two independent fields $h_{\mu\nu}$ and $B_{\alpha_1..\alpha_{d-1},\mu}$ and hence the equation (\ref{4}) should be accompanied by a second
equation
\be
\epsilon_{\alpha_1..\alpha_{d-2}}^{\ \ \ \ \ \ \ \ \rho\sigma}\nabla_\rho h_{\sigma \mu}=m_2 B_{\alpha_1..\alpha_{d-2}, \mu}\, .
\lb{5}
\ee
Equations (\ref{4}) and (\ref{5}) demonstrate a certain duality between fields $h_{\mu\nu}$ and $B_{\alpha_1..\alpha_{d-1},\mu}$.
Only in three dimensions, $d=3$, the tensor $B$ has two indexes and can be identified with the tensor $h_{\mu\nu}$ so that in this case we have a self-dual description (\ref{1}) of a massive graviton. 

In the present paper we mostly focus on the four-dimensional case, $d=4$, and study how the equations of the type (\ref{4}) and (\ref{5}) can be obtained
from an action principle. Since the peculiarities of the appropriate action principle can already  be seen in  three dimensions we start by reviewing the known results in $d=3$.
Throughout the paper the background metric $g_{\mu\nu}$ is considered to be a maximally symmetric metric, so that one has for the Riemann tensor in $d$ dimensions 
\be
R^{\alpha\beta}_{\ \ \mu\nu}=\frac{R}{d(d-1)}(\delta^\alpha_\mu \delta^\beta_\nu-\delta^\alpha_\nu\delta^\beta_\mu)\, ,
\lb{6}
\ee
 where the Ricci scalar is constant of any type $(R>0,\, R=0,\, R<0)$.  We also quote the contraction formula for the product of two $\epsilon$-tensors in $d$ dimensions
 \be
 \epsilon^{\mu_1..\mu_p\alpha_1..\alpha_{d-p}}\epsilon_{\mu_1..\mu_p\beta_1..\beta_{d-p}}=(-1)^sp!(d-p)!\delta^{[\alpha_1}_{\beta_1}..\delta^{\alpha_{d-p}]}_{\beta_{d-p}}\, ,
 \lb{epsilon}
 \ee
 where $s$ is the signature of the spacetime, which will be othen used in the paper.

\section{Self-dual massive gravity in three dimensions}
A curious fact about the equation (\ref{1}) is that it can not be obtained from an action written in terms of a symmetric tensor $h_{\mu\nu}$. So that in the action one has to consider a non-symmetric tensor field $h_{\mu\nu}$. The symmetry condition then arises as a consequence of the field equations. The action then takes the form \cite{Aragone:1986hm}
\be
W[h]=\int_{M^3} \frac{1}{2}\left( h_{\alpha\beta}\epsilon^{\alpha\mu\nu}\nabla_\mu h_\nu^{\ \beta}-m(h_{\mu\nu}h^{\nu\mu}-h^2)\right)\, ,
\lb{7}
\ee
where $h=g^{\mu\nu}h_{\mu\nu}$ is the trace. Variation with respect to $h_{\mu\nu}$ gives the equation \cite{Aragone:1986hm}
\be
\epsilon_\mu^{\ \alpha\beta}\nabla_\alpha h_{\beta\nu}=m(h_{\nu\mu}-g_{\mu\nu}h)\, .
\lb{8}
\ee
The equation (\ref{1}) is a consequence of this equation that can be shown in few steps.

First, let us take the covariant divergence of (\ref{8}). Then,  commuting the covariant derivatives and using that for a maximally symmetric metric the Riemann tensor takes the form (\ref{6}), we obtain that
\be
m\nabla^\mu(h_{\nu\mu}-g_{\mu\nu}h)=-\frac{R}{6}\epsilon_\nu^{\ \alpha\beta} h_{\alpha\beta}\, .
\lb{9}
\ee
On the other hand, the contraction of (\ref{8}) with the $\epsilon$-tensor gives
\be
\nabla^\beta(h_{\nu\beta}-g_{\nu\beta}h)=(-1)^sm\epsilon_\nu^{\ \alpha\beta}h_{\alpha\beta}\, ,
\lb{10}
\ee
where an identity $\epsilon^{\alpha\beta\mu}\epsilon_{\mu\sigma\rho}=(-1)^s(\delta^\alpha_\sigma \delta^\beta_\rho-\delta^\alpha_\rho \delta^\beta_\sigma)$ (see (\ref{epsilon})) is used.
Combining these two equations and assuming that the curvature $R$ is generic so that $R\neq (-1)^s6m^2$, we obtain that $h_{\mu\nu}$ satisfies equations
\be
\nabla^\mu(h_{\nu\mu}-g_{\mu\nu}h)=0\, , \ \ \epsilon_\nu^{\ \alpha\beta} h_{\alpha\beta}=0\, .
\lb{11}
\ee
The second equation in (\ref{11}) means that the antisymmetric part of $h_{\mu\nu}$ is zero, i.e. $h_{[\mu\nu]}=0$. Hence, the tensor field $h_{\mu\nu}$ is symmetric.
Moreover, the trace of $h_{\mu\nu}$ vanishes, $h=0$, as follows from  the trace of (\ref{8}) provided the second equation (\ref{11}) is used. Thus, we see that the tensor field $h_{\mu\nu}$ satisfying the equation (\ref{8}) is indeed a symmetric,  traceless and transverse tensor (\ref{2}) and the equation (\ref{8}) reduces to the equation (\ref{1}) in the Introduction.

\section{The linearized massive gravity in four dimensions}
Our goal in this section is to find a generalization of the first order theory in three dimensions to the four-dimensional case.
As we discussed in the Introduction the theory in this case should contain an additional rank-3 tensor field $B_{\alpha\beta,\mu}=-B_{\beta\alpha,\mu}$
 so that one should have two independent 
equations sufficient to prescribe the dynamics for all fields in question. The field equations are of the type (\ref{4}) and (\ref{5}). We start our analysis with derivation of  the action.

\medskip

\noindent{\it 1. The action and the field equations.}  As in three dimensions, in the action we have to assume that the tensor field $h_{\mu\nu}$ is not symmetric,
\be
W[h,B]=\int_{M^4}\left(\frac{m_1}{2}(h_{\mu\nu}h^{\nu\mu}-h^2)+\frac{m_2}{2}(B_{\alpha\beta,\sigma}B^{\alpha\beta,\sigma}-2B^{\alpha}_{\ \beta,\alpha}B^{\sigma\beta}_{\ \ ,\sigma})+
B_{\alpha\beta,}^{\ \ \ \mu}\nabla_{\rho}h_{\mu\sigma}\epsilon^{\sigma\rho\alpha\beta}\right)\, 
\lb{12}
\ee
The equations of motion for the fields $h_{\mu\nu}$ and $B_{\mu\nu,\alpha}$ take the form
\be
&&m_1(h_{\mu\nu}-hg_{\mu\nu})=\epsilon_{\mu\sigma}^{\ \ \alpha\beta}\nabla^\sigma B_{\alpha\beta,\nu}
\, , \lb{13.1} \\
&&m_2(B_{\mu\nu,\alpha}-B_{\mu}g_{\alpha\nu}+B_{\nu}g_{\alpha\mu})=\epsilon_{\mu\nu}^{\ \ \rho\sigma}\nabla_\rho h_{\alpha\sigma}\, ,
\lb{13}
\ee
where we introduced $B_\mu=B_{\mu\sigma,}^{\ \ \ \sigma}$.
These equations describe a massive spin-2 particle as we now show.

\medskip

\noindent{\it 2. Constraints\footnote{In this paper we call ``constraint'' a relation, at most of first order in derivative, which involves only one field, either $h_{\mu\nu}$ or $B_{\mu\nu,\alpha}$, but not both.}.} The covariant divergence of the first equation in (\ref{13}), taking into account that the background metric is maximally symmetric and hence the relation (\ref{6}) should be used,  produces
\be
m_1\nabla^\mu (h_{\mu\nu}-g_{\mu\nu}h)=\frac{R}{12}\epsilon_\nu^{\ \, \alpha\beta\sigma}B_{\alpha\beta,\sigma}
\lb{14}
\ee
while the contraction of the second equation in (\ref{13}) with $\epsilon$-tensor gives
\be
m_2\epsilon_\nu^{\ \, \alpha\beta\sigma}B_{\alpha\beta,\sigma}=-(-1)^s2\nabla^\mu (h_{\mu\nu}-g_{\mu\nu}h)\, .
\lb{15}
\ee
Provided the background curvature $R\neq - 6(-1)^sm_1m_2$ these two equations produce the constraints
\be
\nabla^\mu(h_{\mu\nu}-g_{\mu\nu}h)=0\, , \ \  \epsilon_\nu^{\ \, \alpha\beta\sigma}B_{\alpha\beta,\sigma}=0\, .
\lb{16}
\ee
The second constraint in (\ref{16}) implies that the tensor field $h_{\mu\nu}$ is traceless, $h=0$, that can be seen by taking trace of the first equation in (\ref{13}).

More constraints can be found by playing with equations (\ref{13}). Contracting the first equation in (\ref{13}) with $\epsilon$-tensor we obtain
\be
m_1\epsilon_{\alpha\beta}^{\ \ \ \mu\nu}h_{\mu\nu}=2(-1)^s(\nabla_\alpha B_{\beta}-\nabla_\beta B_\alpha+\nabla^\sigma B_{\alpha\beta,\sigma})
\lb{17}
\ee
while taking the divergence with respect to third index in the equation for $B_{\mu\nu,\alpha}$ in (\ref{13}) we find
\be
m_2( \nabla^\sigma B_{\alpha\beta,\sigma}-\nabla_\beta B_\alpha+\nabla_\alpha B_\beta)=\frac{R}{6}\epsilon_{\alpha\beta}^{\ \ \ \rho\sigma}h_{\rho\sigma}\, .
 \lb{18}
 \ee
 Combining these two equations we obtain the constraints
 \be
\epsilon_{\alpha\beta}^{\ \ \ \mu\nu}h_{\mu\nu}=0\, , \ \ \  \nabla^\sigma B_{\alpha\beta,\sigma}-\nabla_\beta B_\alpha+\nabla_\alpha B_\beta=0\, .
\lb{19}
\ee
The first constraint implies that the tensor field $h_{\mu\nu}$ is symmetric, $h_{[\mu\nu]}=0$. Then, contracting any two indexes in the second equation in (\ref{13})
we conclude that $B_{\alpha}=0$ and, as follows from (\ref{19}), $\nabla^\sigma B_{\mu\nu,\sigma}=0$. One obtains one more constraint  by taking the divergence with respect to first index in  the equation for $B_{\mu\nu,\alpha}$.  By using the relation (\ref{6}) for the Riemann curvature one then finds that $\nabla^\mu B_{\mu\nu,\alpha}=0$.

\medskip

\noindent{\it 3. Number of degrees of freedom.} Let us list all the constraints we have found. The field $h_{\mu\nu}$ satisfies conditions
\be
h_{\mu\nu}=h_{\nu\mu}\, , \ \ h=0\, , \ \ \nabla^\mu h_{\mu\nu}=0\, ,
\lb{21}
\ee
which indicate that the $h_{\mu\nu}$ is a symmetric transverse-traceless tensor. In four dimensions this tensor has 5 independent components, the number of degrees of freedom of a spin-2 particle. The rank-3 tensor field $B_{\mu\nu,\alpha}$ a priori has 24 components. The constraints  
\be
&&B_{\mu\nu,}^{\ \ \ \mu}=0\, , \ \ \ \epsilon^{\mu\nu\alpha\beta}B_{\nu\alpha,\beta}=0\, , \nonumber \\
&&\nabla^\alpha B_{\mu\nu,\alpha}=0\, , \ \ \ \nabla^\mu B_{\mu\nu,\alpha}=0\, 
\lb{22}
\ee
impose 19 conditions\footnote{This can be easily seen in flat Minkowski spacetime by first representing the components  in the form $B_{\mu\nu,\alpha}(k)e^{ik_\sigma x^\sigma}$.  
In the rest frame one has
$k_0\neq 0$ and $k_i=0$, $i=1,2,3$. So that equations (\ref{22}) reduce to conditions: $B_{0i,0}=0$ (3 conditions), $B_{ij,0}=0$ (3 conditions), $B_{0i,j}=0$ (9 conditions), $B_{ij,}^{\ \ \ i}=0$ (3 conditions), $B_{[ij,k]}=0$ (1 condition). }
on the components thus leaving us with  5 independent degrees of freedom, same number as for the tensor field $h_{\mu\nu}$. 
Let us note that in the theory with action (\ref{12}) the fields $h_{\mu\nu}$ and $B_{\mu\nu,\alpha}$ are not independent variables, they are expressed one through the other. 
Provided  $h_{\mu\nu}$ is considered as the primary field then the tensor $B_{\mu\nu,\alpha}$ is uniquely determined by $h_{\mu\nu}$ or vice versa.

\medskip

\noindent{\it 4. Klein-Gordon type massive field equations.} 
With all these constraints the equations (\ref{13}) take the form
\be
m_1h_{\mu\nu}=\epsilon_{\mu\sigma}^{\ \ \ \alpha\beta}\nabla^\sigma B_{\alpha\beta,\nu}\, , \ \
m_2B_{\mu\nu,\alpha}=\epsilon_{\mu\nu}^{\ \ \ \rho\sigma}\nabla_\rho h_{\alpha\sigma}\, ,
\lb{20}
\ee
as announced in eq. (\ref{4}) and (\ref{5}) in the Introduction. 
By squaring the equations (\ref{20}) we arrive at the field equations quadratic in derivatives
\be
(\Box-\frac{R}{3}-m^2)h_{\mu\nu}=0\, , \ \ (\Box-\frac{5R}{12}-m^2)B_{\mu\nu,\alpha}=0\, ,
\lb{23}
\ee
where $\Box=\nabla^\alpha \nabla_\alpha$, $m^2=-(-1)^sm_1m_2/2$ and we used the constraints (\ref{21}), (\ref{22}).

\medskip

\noindent{\it 5. Coupling to matter sources and relation to the torsion.} Let us consider the coupling of the fields $h_{\mu\nu}$ and $B_{\mu\nu,\alpha}$ with matter sources
$t_{\mu\nu}$ and $S_{\mu\nu,\alpha}$ respectively, where $t_{\mu\nu}$ represents a canonical stress-energy tensor of the source and $S_{\mu\nu,\alpha}$ is the spin tensor of the matter source.
The spin tensor is antisymmetric, $S_{\mu\nu,\alpha}=-S_{\nu\mu,\alpha}$. On the other hand, in the presence of the spin tensor the stress-energy tensor is not symmetric.
In a theory of gravity in which the gravitational variables are the metric $g_{\mu\nu}$ and the torsion $Q_{\mu\nu,\alpha}$ the sources satisfy the identities (see \cite{Hehl:1976kj} for more details)
\be
&&t_{[\mu\nu]}+(\tilde{\nabla}^\alpha-2Q^{\alpha})S_{\mu\nu,\alpha}=0\, , \nonumber \\
&&(\tilde{\nabla}^\mu-2Q^\mu)t_{\mu\nu}-2Q_{\mu\nu,}^{\ \ \ \alpha}t^\mu_{\ \alpha}-S^{\alpha\beta,\mu}\tilde{R}_{\alpha\beta\mu\nu}=0\, ,
\lb{**}
\ee
where $Q_\mu=Q_{\mu\nu,}^{\ \ \ \nu}$ and $\tilde{\nabla}_\alpha$ and $\tilde{R}_{\alpha\beta\mu\nu}$ are  respectively the covariant derivative and the curvature defined  with respect to the Riemann-Cartan connection.

The total action then reads
\be
W[h,B,t,S]=W[h,B]+\int_{M^4} \left(t_{\mu\nu}h^{\nu\mu}+ m_2\frac{(-1)^s}{2}\epsilon_{\mu\nu}^{\ \ \ \sigma\rho}S_{\sigma\rho,\alpha} B^{\mu\nu,\alpha}\right)\, .
\lb{24}
\ee
The reasons for the chosen form of the spin-field $B$  coupling will be clear in a moment. In the presence of matter the  field equations read
\be
&&m_1(h_{\mu\nu}-hg_{\mu\nu})-\epsilon_{\mu\sigma}^{\ \ \alpha\beta}\nabla^\sigma B_{\alpha\beta,\nu}+t_{\mu\nu}=0
\, , \nonumber \\
&&m_2(B_{\mu\nu,\alpha}-B_{\mu}g_{\alpha\nu}+B_{\nu}g_{\alpha\mu})-\epsilon_{\mu\nu}^{\ \ \rho\sigma}\nabla_\rho h_{\alpha\sigma}+m_2\frac{(-1)^s}{2}\epsilon_{\mu\nu}^{\ \ \ \sigma\rho}S_{\sigma\rho,\alpha}=0\, .
\lb{25}
\ee
We do not expect all constraints (\ref{21}) and (\ref{22}) to be valid when the coupling to matter is considered. However, we do want the relation 
\be
\nabla^\mu(h_{\mu\nu}-g_{\mu\nu}h)=0
\lb{*}
\ee
to still hold and we want the field $h_{\mu\nu}$ to be symmetric. These two conditions will impose certain restrictions on the stress-energy tensor $t_{\mu\nu}$ and the spin tensor $S_{\mu\nu,\alpha}$ of the matter source. 
Here we shall identify those restrictions. The strategy remains  the same as before. We find from the first equation (\ref{25}) that
\be
m_1\nabla^\mu(h_{\mu\nu}-hg_{\mu\nu})-\frac{R}{12}\epsilon_\nu^{\ \ \alpha\beta\rho}B_{\alpha\beta,\rho}+\nabla^\mu t_{\mu\nu}=0\, .
\lb{26}
\ee
On the other hand, the contraction of second equation in (\ref{25}) with the $\epsilon$-tensor will produce the relation
\be
m_2\epsilon_\nu^{\ \, \alpha\beta\sigma}B_{\alpha\beta,\sigma}+2(-1)^s\nabla^\mu (h_{\mu\nu}-g_{\mu\nu}h)-2m_2S^\rho_{\ \nu,\rho}=0\, .
\lb{27}
\ee
Combining the two equations (\ref{26}) and (\ref{27} and imposing condition (\ref{*}) we find a relation to be satisfied by the matter tensors
\be
\nabla^\mu t_{\mu\nu}=\frac{R}{6}S^\rho_{\ \ \nu,\rho}
\, .
\lb{28}
\ee
In order to analyze the symmetry condition for the tensor $h_{\mu\nu}$ we contract the first equation (\ref{25}) with the $\epsilon$-tensor
and obtain that
\be
m_1\epsilon_{\alpha\beta}^{\ \ \ \mu\nu}h_{\mu\nu}-2(-1)^s(\nabla_\alpha B_{\beta}-\nabla_\beta B_\alpha+\nabla^\sigma B_{\alpha\beta,\sigma})+\epsilon_{\alpha\beta}^{\ \ \ \mu\nu}t_{\mu\nu}=0\, .
\lb{29}
\ee
The we compute the divergence of the second equation (\ref{25}) and find that
\be
m_2( \nabla^\sigma B_{\alpha\beta,\sigma}-\nabla_\beta B_\alpha+\nabla_\alpha B_\beta)-\frac{R}{6}\epsilon_{\alpha\beta}^{\ \ \ \rho\sigma}h_{\rho\sigma}+m_2\frac{(-1)^s}{2}\epsilon_{\alpha\beta}^{\ \ \ \mu\nu}\nabla^\sigma S_{\mu\nu,\sigma}=0\, .
 \lb{30}
\ee
Combining the two equations (\ref{29}) and (\ref{30}) and imposing the symmetry condition  $h_{[\mu\nu]}=0$ we find another relation to be satisfied by the matter source
\be
t_{[\mu\nu]}=\nabla^\sigma S_{\mu\nu,\sigma}\, .
\lb{31}
\ee
It is now not difficult to see that, provided the background torsion is zero and the background metric is maximally symmetric, the equations
(\ref{28}) and (\ref{31}) are identical to the relations (\ref{**}) that appear in the Riemann-Cartan geometry. This in particular explains our choice for the spin coupling in the action
(\ref{24}). Moreover, since in the Riemann-Cartan geometry the spin tensor couples to torsion we can identify the relation between our field $B_{\mu\nu,\alpha}$ and the torsion
$Q_{\mu\nu,\alpha}$:
\be
Q_{\mu\nu,\alpha}=\frac{(-1)^s}{2}\epsilon_{\mu\nu}^{\ \ \ \sigma\rho}B_{\sigma\rho,\alpha}\, .
\lb{32}
\ee
This is a rather surprising observation since we did not implement in the theory any new geometric structure  other than the standard Riemann geometry.
What is interesting and perhaps non-standard from the point of view of the Riemann-Cartan geometry is that the torsion (\ref{32}),  on-shell, is completely determined by the metric
and vice versa. In the standard approach the metric and the torsion are considered as two independent variables.

Having identified our field $B_{\mu\nu,\alpha}$ as the torsion we  can now invert the logic. Let us assume that the matter source satisfies the conditions   (\ref{28}), (\ref{31}) or, equivalently, (\ref{**}). Then we deduce from the field equations (\ref{25}) that 
the field $h_{\mu\nu}$ is symmetric and satisfies the condition (\ref{*}).

\medskip

\noindent{\it 6. The theory expressed in terms of $h_{\mu\nu}$.}  As we have already noted, the equation (\ref{13}) can be used to express the field $B_{\mu\nu,\alpha}$ is terms of the
rank-2 tensor field $h_{\mu\nu}$. After  substitution  back to the action (\ref{12}) this would give us an action, quadratic in derivatives,  expressed in terms of the field $h_{\mu\nu}$ only,
\be
&&W[h]=\int_{M^4}[\frac{m_1}{2}(h_{\mu\nu}h^{\nu\mu}-h^2)+\frac{(-1)^s}{2m_2}\nabla_\rho h_{\nu\sigma}(-\nabla^\rho h^{\nu\sigma}-\nabla^\nu h^{\rho\sigma}\nonumber \\
&&+\nabla^\nu h^{\sigma\rho}-\nabla^\rho h^{\sigma\nu}+\nabla^\sigma h^{\rho\nu}+\nabla^\sigma h^{\nu\rho})]\, .
\lb{33}
\ee
The field equations which follow from this action can be brought to the form
\be
&&{\cal D}_{\mu\nu}(h)=m^2(h_{\mu\nu}-g_{\mu\nu}h)\, , \lb{34} \\
&&{\cal D}_{\mu\nu}(h)=\Box h_{\nu\mu}+\Box h_{\mu\nu}+\nabla^\rho\nabla_\nu h_{\rho\mu}-\nabla^\rho\nabla_\nu h_{\mu\rho}-\nabla^\rho\nabla_\mu h_{\rho\nu}-\nabla^\rho\nabla_\mu h_{\nu\rho}\, , \nonumber 
\ee 
where $m^2=-(-1)^sm_1 m_2/2$.
On a maximally symmetric background the tensor ${\cal D}_{\mu\nu}(h)$ has the following properties
\be
\nabla^\mu {\cal D}_{\mu\nu}(h)=\frac{R}{6}(\nabla^\mu h_{\mu\nu}-\nabla_\nu h)\, , \ \ \ g^{\mu\nu}{\cal D}_{\mu\nu}(h)=-2\nabla^\nu (\nabla^\mu h_{\mu\nu}-\nabla_\nu h)\, .
\lb{35}
\ee
Combining these properties with the equation (\ref{34}) one finds that $\nabla^\mu h_{\mu\nu}=0$ and $h=0$. The tensor ${\cal D}_{\mu\nu}(h)$ then can be brought to the form
\be
{\cal D}_{\mu\nu}(h)=\Box (h_{\mu\nu}+h_{\nu\mu})-\frac{R}{2}h_{\mu\nu}-\frac{R}{6}h_{\nu\mu}-\nabla_\nu \nabla^\rho h_{\mu\rho}-\nabla_\mu \nabla^\rho h_{\nu\rho}\, .
\lb{36}
\ee
The antisymmetric part of the equation (\ref{34}) then reduces to an algebraic equation on the antisymmetric part of $h_{\mu\nu}$
\be
(\frac{R}{6}-m^2)h_{[\mu\nu]}=0\, ,
\lb{37}
\ee
which in a generic case, when $R\neq 6m^2$, implies that the antisymmetric part is vanishing, $h_{[\mu\nu]}=0$.  The equation (\ref{34}) for the symmetric part $h_{(\mu\nu)}$ then reduces to the massive Klein-Gordon equation (\ref{23}). 

Thus, the action (\ref{33}) describes correctly the spin-2 degrees of freedom. It is surprising that this action is different from that of Fierz-Pauli \cite{Fierz:1939ix}. The two actions are different even when considered on a symmetric tensor field $h_{\mu\nu}$. The difference appears in the structure of the kinetic terms. The most striking peculiarity  of the action (\ref{33}) is that it contains a coupling in the kinetic term between the symmetric and antisymmetric parts of the field $h_{\mu\nu}$. Moreover,  the trace $h$ does not appear at all in the kinetic term. The respective  term in the field equation  (\ref{34}) can not be identified with a linearized  expression for a curvature tensor satisfying the Jacobi identity.
Nevertheless, the tensor ${\cal D}_{\mu\nu}(h)$, containing only the second derivatives of $h_{\mu\nu}$, is
divergence-free in flat space-time, $\partial^\mu  {\cal D}_{\mu\nu}(h)=0$. This is a manifestation of the invariance of the kinetic term in (\ref{33}) (and, in fact, of the kinetic term in (\ref{12})) under the gauge symmetry
\be
h_{\mu\nu}\rightarrow h_{\mu\nu}+\partial_\nu\xi_\mu\, ,
\lb{38}
\ee
where $\xi_\mu$ is an arbitrary vector.

\medskip

\noindent{\it 7. The  propagator.}  Let us focus on flat Minkowski spacetime equipped with metric $\eta^{\alpha\beta}$ . The field equation (\ref{34}) for the tensor field $h_{\mu\nu}$ 
is represented as 
\be
{\cal O}^{\nu\sigma\alpha\beta}h_{\alpha\beta}=0\, ,
\lb{38-0}\ee
where the field operator
\be
{\cal  O}^{\nu\sigma\alpha\beta}&=&\frac{1}{2}\left(\Box (\eta^{\alpha\nu}\eta^{\sigma\beta}+\eta^{\alpha\sigma}\eta^{\beta\nu})+\p^\alpha\p^\nu\eta^{\sigma\beta}
-\p^\beta\p^\nu\eta^{\alpha\sigma}-\p^\alpha\p^\sigma\eta^{\nu\beta}-\p^\beta\p^\sigma\eta^{\alpha\nu}\right)\nonumber \\
&-&m^2\pl\eta^{\alpha\sigma}\eta^{\beta\nu}
-\eta^{\nu\sigma}\eta^{\alpha\beta}\pr\, .
\lb{38-1}
\ee
In the momentum space we have to replace $\partial_\alpha\rightarrow ik_\alpha$. The propagator ${\cal P}_{\alpha\beta\mu\nu}$ then  satisfies the relation (we remind that the field $h_{\mu\nu}$ is not a priori symmetric)
\be
{\cal P}_{\alpha\beta\mu\nu}{\cal O}^{\mu\nu\sigma\rho}=\delta^\sigma_\alpha \delta^\rho_{\beta}\, .
\lb{38-2}
\ee
The propagator can be decomposed on symmetric and antisymmetric parts with respect to the two groups of indexes,
\be
{\cal P}_{\alpha\beta\mu\nu}={\cal P}_{(\alpha\beta)(\mu\nu)}+{\cal P}_{[\alpha\beta](\mu\nu)}+{\cal P}_{(\alpha\beta)[\mu\nu]}+{\cal P}_{[\alpha\beta][\mu\nu]}\, .
\lb{38-3}
\ee
For the symmetric components of the propagator we find 
\be
{\cal P}_{(\alpha\beta)(\mu\nu)}={\cal P}^{PF}_{\alpha\beta\mu\nu}+\frac{1}{4m^4}(\eta_{\beta\nu}k_\alpha k_\mu+\eta_{\alpha\nu}k_\beta k_\mu+\eta_{\beta\mu}k_\alpha k_\nu+\eta_{\alpha\mu}k_\beta k_\nu)\, ,
\lb{38-4}
\ee
where ${\cal P}^{PF}_{\alpha\beta\mu\nu}$ is the propagator in the Fierz-Pauli theory \cite{Fierz:1939ix} of massive gravity,
\be
{\cal P}^{PF}_{\alpha\beta\mu\nu}=-\frac{1}{k^2+m^2}\left(\frac{1}{2}(P_{\alpha\mu}P_{\beta\nu}+P_{\alpha\nu}P_{\beta\mu})-\frac{1}{3}P_{\alpha\beta}P_{\mu\nu}\right)\, ,
\lb{38-41}
\ee
where $P_{\alpha\beta}=\eta_{\alpha\beta}-\frac{k_\alpha k_\beta}{m^2}$. For the  other components of the propagator we find that
\be
{\cal P}_{[\alpha\beta](\mu\nu)}={\cal P}_{(\mu\nu)[\alpha\beta]}=-\frac{1}{4m^4}(\eta_{\mu\beta}k_\nu k_\alpha-\eta_{\mu\alpha}k_\nu k_\beta +\eta_{\nu\beta}k_\mu k_\alpha-
\eta_{\nu\alpha}k_\mu k_\beta )\, ,
\lb{38-5}
\ee
\be
{\cal P}_{[\alpha\beta][\mu\nu]}=-\frac{1}{2m^2} \left(
 \eta_{\mu\alpha}\eta_{\nu\beta}-\eta_{\mu\beta}\eta_{\nu\alpha}-\frac{1}{2m^2}\crl \eta_{\nu\alpha} k_{\mu}k_{\beta}
+\eta_{\mu\beta}k_\nu k_\alpha-\eta_{\nu\beta} k_\mu k_\alpha-\eta_{\mu\alpha} k_\nu k_\beta\crr\right) \, .
\lb{38-6}
\ee
We see that the propagator (\ref{38-3}) in the theory (\ref{34}) differs from the propagator in the Fierz-Pauli theory only  by regular terms. Thus, the only singularity in the propagator is identical to that of the Fierz-Pauli theory that corresponds to a massive spin-2 particle. The theory (\ref{34}) then does not contain any  physical or non-physical states (considered as poles in the propagator) other than that of the massive graviton. We should note that the theory (\ref{34}) could be viewed as  a counter-example to the statement made long ago in \cite{VanNieuwenhuizen:1973fi} that the only consistent theory of a rank-2 tensor $h_{\mu\nu}$ that does not  contain neither ghosts nor tachyons  is the one in which the symmetric and antisymmetric parts of $h_{\mu\nu}$ decouple. The theory at hand does  contain this kind of coupling, nevertheless it is free of the mentioned  pathologies. 
The antisymmetric part of $h_{\mu\nu}$ does not seem to be dynamical in the theory (\ref{33}) since the relevant part (\ref{38-6}) of propagator  does not contain any pole, although its presence is essential to have a consistent theory. In fact, imposing the condition that the antisymmetric part of $h_{\mu\nu}$ vanishes in the action (\ref{33}) we would get a theory of the symmetric $h_{\mu\nu}$ that contains ghosts. Similarly, imposing the condition that 
the symmetric part of $h_{\mu\nu}$  is zero in (\ref{33}) we would get a theory that effectively describes a tachyonic spin-1 massive mode. Only when the both parts, symmetric and antisymmetric, of $h_{\mu\nu}$ are present in the action we have a fully consistent theory without ghosts and/or tachyons. In Appendix we give some details of this analysis.

\medskip

\noindent{\it 8. The theory expressed in terms of torsion $Q_{\mu\nu,\alpha}$.} On the other hand, we can use the equation (\ref{13.1}) and express the field $h_{\mu\nu}$ in terms of $B_{\mu\nu,\alpha}$ or, using the relation (\ref{32}), in terms of the torsion tensor $Q_{\mu\nu,\alpha}$. The theory then is formulated entirely in terms of the torsion with the quadratic action in the form
\be
W[Q]=-\frac{2}{m_1}\int_{M^4}\left(H_{\mu\nu}H^{\nu\mu}-\frac{1}{3}H^2+m^2Q_{\mu\nu,\alpha}Q^{\mu\alpha,\nu}\right)\, ,
\lb{39}
\ee
where $m^2=-(-1)^sm_1m_2/2$ as before and we introduced $H_{\mu\nu}=\nabla^\sigma Q_{\mu\sigma,\nu}$, $H=g^{\mu\nu}H_{\mu\nu}$. The field equation satisfied by $Q_{\mu\nu,\alpha}$ is
\be
\nabla_\mu H_{\alpha\nu}-\nabla_\nu H_{\alpha\mu}+\frac{1}{3}(g_{\mu\alpha}\nabla_\nu H-g_{\nu\alpha}\nabla_\mu H)+m^2(Q_{\mu\alpha,\nu}-Q_{\nu\alpha,\mu})=0\, .
\lb{40}
\ee
This constitutes a formulation of the linearized massive gravity  that is dual to the metric formulation in terms of rank-2 tensor field as given by equations (\ref{33}), (\ref{34}). 

Let us analyze the equations (\ref{40}).  For simplicity we consider flat Minkowski spacetime as a background. In flat spacetime we have that
$\partial^\mu H_{\mu\nu}=\partial^\mu\partial^\sigma Q_{\mu\sigma, \nu}=0$ due to the antisymmetric property of the torsion tensor, $Q_{\mu\sigma,\nu}=-Q_{\sigma\mu,\nu}$.
Now, taking the divergence $\partial^\alpha$ of (\ref{40}) we obtain that $\partial^\alpha Q_{\mu\alpha,\nu}=\partial^\alpha Q_{\nu\alpha,\mu}$. This indicates that the earlier introduced tensor $H_{\mu\nu}=\partial^\alpha Q_{\mu\alpha,\nu}$ is symmetric, $H_{\mu\nu}=H_{\nu\mu}$. As we will see in a moment namely this tensor will contain the degrees of freedom
of a spin-2 particle. 

The one more divergence, this time with respect to index $\mu$, of equation (\ref{40}) results in the equation
\be
\Box H_{\alpha\nu}+\frac{1}{3}(\partial_\alpha \partial_\nu H-g_{\nu\alpha}\Box H)-m^2H_{\alpha\nu}-m^2\partial^\mu Q_{\nu\alpha,\mu}=0\, .
\lb{41}
\ee
The antisymmetric part of this equation gives the constraint $\partial^\mu Q_{\nu\alpha,\mu}=0$ while the symmetric part takes the form of an   equation for tensor  $H_{\mu\nu}$
\be
\Box H_{\alpha\nu}+\frac{1}{3}(\partial_\alpha \partial_\nu H-g_{\nu\alpha}\Box H)-m^2H_{\alpha\nu}=0\, .
\lb{42}
\ee
The trace of this equation results in the condition that $H=0$ so that (\ref{42}) reduces to the Klein-Gordon massive equation for $H_{\mu\nu}$. Collecting all the equations  obtained for the tensor $H_{\mu\nu}$  we find that
\be
&&H_{\mu\nu}=H_{\nu\mu}\, , \  \ \partial^\mu H_{\mu\nu}=0 \, \ \ g^{\mu\nu}H_{\mu\nu}=0\, , \nonumber \\
&&(\Box-m^2)H_{\mu\nu}=0\, .
\lb{43}
\ee
These are exactly the conditions to be satisfied by a symmetric tensor field which describes a spin-2 particle. However, in this construction the field $H_{\mu\nu}$ is not 
a primary object but rather it is built from the torsion tensor. 

Since the trace $H=0$ the equation (\ref{40}) then takes the form
\be
\partial_\mu H_{\alpha\nu}-\partial_\nu H_{\alpha\mu}+m^2(Q_{\mu\alpha,\nu}-Q_{\nu\alpha,\mu})=0\, .
\lb{44}
\ee
It can be used to express the torsion tensor in terms of the symmetric tensor $H_{\mu\nu}$ satisfying equations (\ref{43}) as follows
\be
Q_{\mu\alpha,\nu}=-\frac{1}{m^2}(\partial_\mu H_{\alpha\nu}-\partial_\alpha H_{\mu\nu})\, .
\lb{45}
\ee
Together  with the condition that $H_{\mu\nu}=\partial^\alpha Q_{\mu\alpha,\nu}$ this constitutes the field equations to be satisfied by the torsion tensor $Q_{\mu\nu,\alpha}$.

Below we list all the constraints imposed by the field equations on the torsion tensor:
\be
&&Q^\alpha_{\ \mu,\alpha}=0\, , \ \ \partial^\mu Q_{\nu\alpha,\mu}=0\, ,  \ \ \partial^\alpha Q_{\mu\alpha,\nu}=\partial^\alpha Q_{\nu\alpha,\mu}\, ,\nonumber \\
&&Q_{\mu\alpha,\nu}+Q_{\alpha\nu,\mu}+Q_{\nu\mu,\alpha}=0\, .
\lb{46}
\ee
The last condition follows directly from the representation (\ref{45}). A more detailed analysis of the constraints can be done using the  the momentum representation for the torison tensor $Q_{\mu\nu,\alpha}(k) e^{ik_\mu x^\mu}$ and choosing the coordinate system so that $k_0=m$ and $k_i=0$, $i=1,2,3$.  Then the analysis of equation (\ref{45})
shows that the only non-vanishing components of the torsion are $Q_{0 i,j}$ subject to the symmetry condition $Q_{0i,j}=Q_{0j,i}$ and the trace condition $\delta^{ij}Q_{0i,j}=0$.
These conditions leave exactly 5 non-vanishing components. These  are the non-vanishing components of the tensor $H_{\mu\nu}=-k_0 Q_{\mu 0,\nu}$ satisfying (\ref{43}).
Finally, let us note that, as follows from  the equation (\ref{20}), the relation between the field $h_{\mu\nu}$ and the torsion is $h_{\mu\nu}=2/m_1 (-1)^s\nabla^\sigma Q_{\mu\sigma,\nu}=2/m_1(-1)^sH_{\mu\nu}$ that indeed identifies the tensor $H_{\mu\nu}$ built from the torsion tensor as the one which describes a spin-2 particle.

It should be noted that our construction based on the action (\ref{39}) is very similar to the recent study made in \cite{Nair:2008yh} where the vierbein
 and  the spin connection as independent gravitational variables are considered.  This theory, additionally to a massless graviton, contains a massive spin-2 propagating mode
 which originates from the torsion tensor.  In particular, the equation (\ref{45}) appears in the linearized equation for the  torsion field in the model of ref.\cite{Nair:2008yh}.
 This suggests that the two theories  may be closely related. 
 We however note that  in the theory we consider in the present paper the ``metric'' $h_{\mu\nu}$ and the torsion are not two independent variables. They are expressed one through the other. Hence, there is only one propagating mode, that of the massive spin-2 particle. Possibly, there is a certain truncation of the model of ref.\cite{Nair:2008yh}
that would give rise to the theory considered here. The other related work is  \cite{Zinoviev:2008ze}.  The model considered in that work involves the vierbein and the torsion (or the Lorentz connection) and, after gauging away the  Stueckelberg fields, it appears to take a form similar 
(but not identical)  to the one considered here. However, as the analysis shows, this theory being formulated in terms of $h_{\mu\nu}$, reduces
to  the Pauli-Fierz theory for the symmetric part of $h_{\mu\nu}$
plus the massive term (without a kinetic term) for the antisymmetric part. There is no
coupling between the symmetric and anti-symmetric parts of $h_{\mu\nu}$ in this theory. We could not find a local transformation which would bring our theory (\ref{33})  to this diagonal form
and we believe that these two theories are not equivalent. The other similar approaches which however lead to the standard Fierz-Pauli theory are \cite{Boulanger:2003vs}.

\medskip

\noindent{\it 9. Generalization to higher dimensions.}  As we discussed in the Introduction, in a higher dimension $d$ the tensor field $h_{\mu\nu}$ should be supplemented with a
rank-(d-1) tensor $B_{\alpha_1..\alpha_{d-2},\mu}$. This tensor is  equivalent to a rank-3 tensor $Q_{\mu\nu,\sigma}\sim \epsilon_{\mu\nu}^{\ \ \ \ \alpha_1..\alpha_{d-2}}B_{\alpha_1..\alpha_{d-2},\sigma}$, where we omit the exact pre-factor, which can still be identified with the torsion.
 The theory then is most easily formulated in terms of fields $h_{\mu\nu}$ and $Q_{\mu\nu,\sigma}$. The universal action which describes a massive spin-2 particle 
 in arbitrary dimension $d$ then takes the form
 \be
 W[h, Q]=\int_{M^d}\left(\frac{m_1}{2}(h_{\mu\nu}h^{\nu\mu}-h^2)+m_2Q_{\mu\nu,\sigma} Q^{\mu\sigma,\nu}+2Q^{\alpha\beta,\mu}\nabla_\beta h_{\mu\alpha}  \right)\, .
\lb{47}
\ee
In four dimensions this action is obtained from (\ref{12}) by re-expressing the field $B$ in terms of the torsion using relation (\ref{32}) and after some re-definition of parameters 
$m_1$ and $m_2$ in order to absorb the signature dependent factor $(-1)^s$. We have checked that in any dimension $d$ this action still describes a transverse-traceless
symmetric field $h_{\mu\nu}$ (the symmetry condition follows from the field equations) which satisfies a Klein-Gordon type equation with the mass $m^2=-m_1m_2/2$.

\section{Conclusions} 

In the current literature on the massive gravity it is believed that, at the linearized level, the only consistent theory
to be used is that of the Fierz-Pauli which is formulated in terms of the symmetric rank-2 tensor. The latter is naturally identified with the
components of the metric (or, more precisely, with the deviation of the curved metric from that of Minkowski spacetime). In the present paper we show that
this description is not unique.  Giving up the ``symmetry condition'', we formulate a theory which contains both the symmetric and antisymmetric
parts of $h_{\mu\nu}$. On the field equations the antisymmetric  part of $h_{\mu\nu}$ vanishes.
The remaining propagating degrees of freedom are that of a massive spin-2 particle.  Contrary to some expectations, this theory is free of possible pathologies (ghosts or tachyons).

On the other hand, even the rank-2 tensor is not obligatory to use when we want to describe a massive spin-2 particle. In the other proposed formulation, which is dual to the one in terms of $h_{\mu\nu}$, the massive spin-2 particle is described entirely in terms of a rank-3 tensor (torsion) $Q_{\mu\nu,\alpha}$. The equivalence between these two formulations is demonstrated
by means of the action which contains both (non-symmetric) $h_{\mu\nu}$ and the torsion $Q_{\mu\nu,\alpha}$ and is linear in derivatives.  These equivalent formulations exist in any
dimension $d\geq 4$.

Of course, the most difficult part in formulating  the massive gravity starts at the  non-linear level when the self-interactions are introduced.
Noting the remarkable recent progress in constructing such a formulation based on the Fierz-Pauli theory we believe that the class of theories introduced in this paper should not be a priori excluded
from the consideration. The study of the non-linear versions of these theories  may lead to interesting and perhaps surprising development.

\bigskip

\noindent{\large \bf Acknowledgements} 

S.S. would like to acknowledge the useful discussions with K. Noui and X. Bekaert on the earlier stage of the project. We like to thank A. Barvinsky, F. Hehl, Yu. Obukhov and R. Metsaev  for useful comments.

\appendix
\section{Appendix}
\setcounter{equation}0
In this section, we give the explicit expression for the propagator associated with the (anti)-symmetric part of $h_{\mu\nu}$ and emphasize the crucial 
role played by the coupling between the symmetric and antisymmetric terms in getting rid of ghosts and tachyons. 
~\\The action (\ref{33}) with  Minkowski  spacetime as a background (with $\mathcal{O}$ defined by (\ref{38-1})) reads
\bea
W[h]=\int_{M^4}\frac{\pl-1\pr^s}{m_2}\crl h_{\alpha\beta}\mathcal{O}^{\alpha\beta\mu\nu}h_{\mu\nu}\crr\, .\label{app1} 
\eea
We write the various symmetrisations of the differential operator $\mathcal{O}$ : 
\bea
\mathcal{O}^{(\alpha\beta)[\mu\nu]}&=&\frac{1}{4}\pl\p^{\alpha}\p^{\mu}\eta^{\nu\beta}+\p^{\beta}\p^{\mu}\eta^{\alpha\nu}
-\p^\alpha\p^\nu\eta^{\mu\beta}-\p^{\beta}\p^{\nu}\eta^{\mu\alpha}\pr \, ,\\
\mathcal{O}^{(\alpha\beta)(\mu\nu)}&=&\frac{1}{2}\Bigg[\Box\eta^{\alpha\mu}\eta^{\nu\beta}+\Box\eta^{\alpha\nu}\eta^{\mu\beta}
-\frac{1}{2}\pl\p^\alpha\p^\nu\eta^{\mu\beta}+\p^{\alpha}\p^{\mu}\eta^{\nu\beta}+\p^{\beta}\p^{\nu}\eta^{\mu\alpha}+\p^{\beta}\p^{\mu}\eta^{\alpha\nu}
\pr\nn\\&&-2m^2\pl\frac{1}{2}\pl\eta^{\alpha\nu}\eta^{\beta\mu}+\eta^{\alpha\mu}\eta^{\beta\nu}\pr-\eta^{\mu\nu}\eta^{\alpha\beta}\pr\Bigg]\, ,\nn\\
\mathcal{O}^{[\alpha\beta][\mu\nu]}&=&\frac{1}{4}\pl\p^{\alpha}\p^{\mu}\eta^{\nu\beta}+
\p^{\beta}\p^{\nu}\eta^{\mu\alpha}-\p^\alpha\p^\nu\eta^{\mu\beta}-\p^{\beta}\p^{\mu}\eta^{\alpha\nu}\pr
-\frac{m^2}{2}\pl\eta^{\alpha\nu}\eta^{\beta\mu}-\eta^{\alpha\mu}\eta^{\beta\nu}\pr\, .\nn
\eea
We note that $\mathcal{O}^{[\mu\nu](\alpha\beta)}=\mathcal{O}^{(\alpha\beta)[\mu\nu]}$ and that 
(in the Fourier formulation) 
$k_{\alpha}k_{\beta}\mathcal{{O}}^{(\alpha\beta)[\mu\nu]}=0$  and is therefore not invertible. 
We now give the expressions for the propagators associated with the totally (anti)-symmetric parts of $\mathcal{O}$ : 
\bea
&&\mathcal{\tilde{P}}_{(\mu\nu)(\alpha\beta)}=\frac{-1}{k^2+m^2}\Bigg[\frac{1}{2}\pl\eta_{\mu\alpha}\eta_{\beta\nu}+\eta_{\mu\beta}\eta_{\alpha\nu}
\pr-\frac{1}{3}\eta_{\mu\nu}\eta_{\alpha\beta}+\frac{1}{2\pl k^2+2m^2\pr}\Big[ k_\mu k_\alpha\eta_{\beta\nu}\\&&
+k_{\mu} k_{\beta}\eta_{\nu\alpha}+ k_\nu k_\alpha\eta_{\mu\beta}+ k_\nu k_\beta\eta_{\alpha\mu}\Big]\nn-\frac{1}{3m^2}\pl k_{\mu} k_{\nu}
\eta_{\alpha\beta}+ k_{\alpha} k_{\beta}\eta_{\mu\nu}\pr-\frac{1}{3}\frac{k^2-m^2}{m^4\pl k^2+2m^2\pr} k_\alpha k_\beta k_\mu k_\nu\Bigg]\nn\\
&&\mathcal{\tilde{P}}_{[\mu\nu][\alpha\beta]}=\frac{1}{2m^2}\crl\pl\eta_{\mu\alpha}\eta_{\beta\nu}-\eta_{\mu\beta}\eta_{\alpha\nu}\pr
+\frac{1}{k^2-2m^2}\pl k_{\mu} k_{\beta}\eta_{\alpha\nu}+ k_{\alpha} k_{\nu}\eta_{\beta\mu}- k_{\mu} k_{\alpha}\eta_{\beta\nu}
- k_{\nu} k_{\beta}\eta_{\alpha\mu}\pr\crr\nn
\eea
These reduced propagators contains the supplementary poles (including tachyonic ones) compared to the one present in the total propagator (\ref{38-3}). We now examine how these pathological poles are cured in the complete theory via the coupling  between $h_{(\mu\nu)}$ and $h_{[\mu\nu]}$. 
Resolving equation (\ref{38-2}) in the complete theory we find that due to the mixing part in the field operator the symmetric part of the propagator $P_{(\alpha\beta)(\mu\nu)}$ is inverse to a new operator 
\bea
\mathcal{O}_{new}^{(\alpha\beta)(\mu\nu)}=\mathcal{O}^{(\alpha\beta)(\mu\nu)}-\mathcal{O}^{(\alpha\beta)[\rho \sigma]}\mathcal{\tilde{P}}_{[\rho\sigma][\lambda\gamma]}\mathcal{O}^{[\lambda\gamma](\mu\nu)}\, 
\lb{sym}
\eea
and the antisymmetric part of the propagator $P_{[\alpha\beta][\mu\nu]}$ is inverse to the operator
\bea
\mathcal{O}_{new}^{[\alpha\beta][\mu\nu]}=\mathcal{O}^{[\alpha\beta][\mu\nu]}-\mathcal{O}^{[\alpha\beta](\rho \sigma)}\mathcal{\tilde{P}}_{(\rho\sigma)(\lambda\gamma)}\mathcal{O}^{(\lambda\gamma)[\mu\nu]}\, .
\lb{asym}
\eea
Although $\mathcal{O}^{[\alpha\beta][\mu\nu]}$ alone gives rise to additional and potentially pathological modes, the operator $\mathcal{O}_{new}^{[\alpha\beta][\mu\nu]}$, 
comprising the mixing term, contains only healthy modes. This can be seen by inverting it which gives back the (manifeslty regular) propagator (\ref{38-6}). The same is true for the
symmetric operator (\ref{sym}) and similarly, the mixing term kills the supplementary modes and the total differential operator acting
 on $h_{(\mu\nu)}$ is healthy and admits (\ref{38-4}) as inverse.  The mixing part of the propagator (\ref{38-5}) then is expressed as follows
 \be
 P_{[\alpha\beta](\mu\nu)}=-P_{[\alpha\beta][\gamma\lambda]}\mathcal{O}^{[\gamma\lambda](\sigma\rho)}\tilde{P}_{(\sigma\rho)(\mu\nu)}\, .
\lb{sym-asym}
\ee
\setcounter{equation}0


\newpage
\begin{thebibliography}{999}

{\frenchspacing \parskip=2mm

\bibitem{Hinterbichler:2011tt} 
 K.~Hinterbichler,
  arXiv:1105.3735 [hep-th].\\
V.~A.~Rubakov and P.~G.~Tinyakov,
  Phys.\ Usp.\  {\bf 51}, 759 (2008).\\
    G.~R.~Dvali, G.~Gabadadze and M.~Porrati,
  Phys.\ Lett.\ B {\bf 485}, 208 (2000)
 .\\
  T.~Damour, I.~I.~Kogan and A.~Papazoglou,
  Phys.\ Rev.\ D {\bf 67} (2003) 064009.\\
   N.~Arkani-Hamed, H.~Georgi and M.~D.~Schwartz,
  Annals Phys.\  {\bf 305}, 96 (2003)
 .\\
    C.~de Rham, G.~Gabadadze and A.~Tolley,
  arXiv:1107.3820 [hep-th];
  Phys.\ Rev.\ Lett.\  {\bf 106}, 231101 (2011);
  Phys.\ Lett.\ B {\bf 711}, 190 (2012).\\
  S.~F.~Hassan and R.~A.~Rosen,
  Phys.\ Rev.\ Lett.\  {\bf 108}, 041101 (2012).\\
   M.~Mirbabayi,
  arXiv:1112.1435 [hep-th].\\
  K.~Hinterbichler and R.~A.~Rosen,
  arXiv:1203.5783 [hep-th].\\
   E.~Babichev, C.~Deffayet and R.~Ziour,
  Phys.\ Rev.\ D {\bf 82}, 104008 (2010)
 ; 
  Phys.\ Rev.\ Lett.\  {\bf 103}, 201102 (2009)
 .\\
   L.~Alberte, A.~H.~Chamseddine and V.~Mukhanov,
    JHEP {\bf 1104}, 004 (2011);
  JHEP {\bf 1012}, 023 (2010)
 .\\
   A.~H.~Chamseddine and V.~Mukhanov,
  JHEP {\bf 1108}, 091 (2011)
  ;  
  JHEP {\bf 1008}, 011 (2010).\\
  S.~F.~Hassan, A.~Schmidt-May and M.~von Strauss,
  arXiv:1203.5283 [hep-th].
  
\bibitem{Aragone:1986hm} 
  C.~Aragone and A.~Khoudeir,
  Phys.\ Lett.\ B {\bf 173}, 141 (1986).
   
\bibitem{Tyutin:1997yn} 
  I.~V.~Tyutin and M.~A.~Vasiliev,
  Teor.\ Mat.\ Fiz.\  {\bf 113N1}, 45 (1997).

\bibitem{Sachs:2008gt}
  W.~Li, W.~Song and A.~Strominger,
  JHEP {\bf 0804}, 082 (2008).\\
  I.~Sachs and S.~N.~Solodukhin,
  JHEP {\bf 0808}, 003 (2008).

\bibitem{Deser:1982vy} 
  S.~Deser, R.~Jackiw and S.~Templeton,
  Phys.\ Rev.\ Lett.\  {\bf 48}, 975 (1982).\\
  S.~Deser, R.~Jackiw and S.~Templeton,
  Annals Phys.\  {\bf 140}, 372 (1982)


 
\bibitem{Hehl:1976kj} 
 A.~Trautman,
  gr-qc/0606062.\\
   F.~W.~Hehl, J.~D.~McCrea, E.~W.~Mielke and Y.~Ne'eman,
  Phys.\ Rept.\  {\bf 258}, 1 (1995).
  F.~W.~Hehl, P.~Von Der Heyde, G.~D.~Kerlick and J.~M.~Nester,
  Rev.\ Mod.\ Phys.\  {\bf 48}, 393 (1976).\\
  Y.~N.~Obukhov, V.~N.~Ponomarev and V.~V.~Zhytnikov,
  Gen.\ Rel.\ Grav.\  {\bf 21} (1989) 1107.
  
\bibitem{Fierz:1939ix}
  M.~Fierz and W.~Pauli,
  Proc.\ Roy.\ Soc.\ Lond.\  A {\bf 173} (1939) 211.

\bibitem{VanNieuwenhuizen:1973fi}
  P.~Van Nieuwenhuizen,
  Nucl.\ Phys.\  B {\bf 60}, 478 (1973).
  
  
\bibitem{Nair:2008yh} 
  V.~P.~Nair, S.~Randjbar-Daemi and V.~Rubakov,
  Phys.\ Rev.\ D {\bf 80}, 104031 (2009).\\
  V.~Nikiforova, S.~Randjbar-Daemi and V.~Rubakov,
  Phys.\ Rev.\ D {\bf 80}, 124050 (2009).\\
 C.~Deffayet and S.~Randjbar-Daemi,
  Phys.\ Rev.\ D {\bf 84}, 044053 (2011).
  
\bibitem{Zinoviev:2008ze}
  Y.~M.~Zinoviev,
  Nucl.\ Phys.\ B {\bf 808} (2009) 185.
\bibitem{Boulanger:2003vs}
 P.~C.~West,
  Class.\ Quant.\ Grav.\  {\bf 18}, 4443 (2001).\\
  N.~Boulanger, S.~Cnockaert and M.~Henneaux,
  JHEP {\bf 0306}, 060 (2003).\\
    B.~Gonzalez, A.~Khoudeir, R.~Montemayor and L.~F.~Urrutia,
  JHEP {\bf 0809}, 058 (2008).\\
  H.~Casini, R.~Montemayor and L.~F.~Urrutia,
  Phys.\ Rev.\ D {\bf 66}, 085018 (2002).\\
  H.~Casini, R.~Montemayor and L.~F.~Urrutia,
  Phys.\ Rev.\ D {\bf 68}, 065011 (2003).\\
} 
 
\end{thebibliography}
\end{document}